# A Simple Solution To The Uncertain Delay Problem in USRP Based SDR-Radar Systems

Andriyan Bayu Suksmono
School of Electrical Engineering and Informatics
Institut Teknologi Bandung, Jl. Ganesha No.10,
Bandung, Indonesia; Email: suksmono@stei.itb.ac.id

**Abstract**

We propose a simple solution to the uncertain delay problem in USRP (Universal Software Radio Peripheral)-based SDR (Software-Defined Radio)-radar systems. Instead of time-synchronization as employed in (pseudo-) passive radar configurations, which require two or more synchronized receivers, we use direct reception (antenna-to-antenna) signal in a single receiver system as a reference to the exact location of the target echoes. After finding the reference position, reordering of the echoes is conducted by circular shift so that the reference moved to the origin. We demonstrate the effectiveness of the proposed method by simulating the problem in Matlab and implementing a 128 length random code radar in a USRP. The random code is constructed from zero padded Barker sequence product. Experiments on measuring multiple echoes of the targets at precise range bins confirm the applicability of the proposed method.

## 1. Introduction

The SDR (Software Defined Radio) [1] offers flexibility in the design and implementation of various radio systems. One of the popular tools for implementing SDR is the free and open source software called GNU Radio, whereas the hardware counterpart is Ettus' USRP (Universal Software Radio Pheripheral). The affordability of the USRP drives creativity among SDR researcher and enthusiasts in implementing a radio system, such as using the GNU Radio and USRP to build a radar, creating a new system called SDR-Radar. The successfulness of this system will make a great impact in radar engineering, since various kind of radar can be implemented in a same hardware.

A few success story emerges in SDR-Radar implementation. Most of them utilize multistatic configuration [2], [3], [4], in which more than one USRP employed in the radar system. The motivation of using multistatic configuration could be understood, considering the ranging in a radar is highly rely on the transmit-receive time delay. On the other hand, the USRP implementation that considers the device as a communication point to a host computer, whether connected by a USB or ethernet cable, introduces uncertainty or randomness of the delay. Since timing is crucial in radar, such randomness will severely affect ranging performance of the SDR-radar. Therefore, a pair of synchronized receiver is required to assure precise timing of the incoming waves.

In this paper, we show that correct delay time measurement can be performed with only a single receiver. The idea is by using direct reception of the signal, usually considered as a drawback on undermining delay from far-range objects of interest, as a reference. We calculate the center of the direct reception, perform correction, and show that the expected echoes will appear at precise position in the range bin. A random code radar is build to demonstrate the workability of the propose system.

The rest of the paper is organized as follows. Section 2 describes the USRP-based radar systems and the random delay problem. The proposed solution to surmount the problem is explained in Section 3, which is tested by experiments given in Section 4. We conclude the paper in Section 5.

## 2. USRP-Based SDR-Radars and Uncertain Delay Problem

There are two main parts in an SDR, namely the software and hardware parts. The first one dealing with signal processing, whereas the later take care of air interface. GNU Radio, which provides signal processing blocks to build a radio system, represents the S/W parts. The USRP, on the other hand, when connected to a host computer running the GNU Radio implement the hardware part of the system. The data transfer between the GNU Radio running in a host PC with USRP utilize USB (Universal Serial Bus) in the USRP1 series, Ethernet in USRP2, or GPIO (General Purpose Input/Output)/ SPI (Serial Peripheral Interface) on the bus series. In the first two series in particular, the communication protocol will generally introduce random delay to the transmitted signal. Since radar rely on range measurement on the delay between transmit and received signal, precise timing is a crucial issue.

Although the random delay problem has not explicitly been addressed in existing USRP-based radar implementations, the success story on USRP-based radars come from synchronized-receiver configuration. In [2], passive bistatic radar has been built using 2 synchronized USRP N210, where UMTS and DVB-T transmitters are used as illuminators. In [3], a passive OTH (Over the Horizon) Radar was built with 6 USRP2 synchronized with GPS clock. In addition, a pseudo passive OFDM radar was also built using 2 USRPs synchronized with MIMO cable [4]. Whereas the first two systems use external source undedicated for the radar, the OFDM radar transmit its own signal using one of the transmitter and receives the echoes at two synchronized USRP receivers, forming a pseudo-passive radar. Although exact transmission and arrival time of the wave are not known, receiver synchronization guarantee the exact timing of the receive wave. Then, cross-correlation between signals arriving at first receiver with the second one will represent a correct delay time.

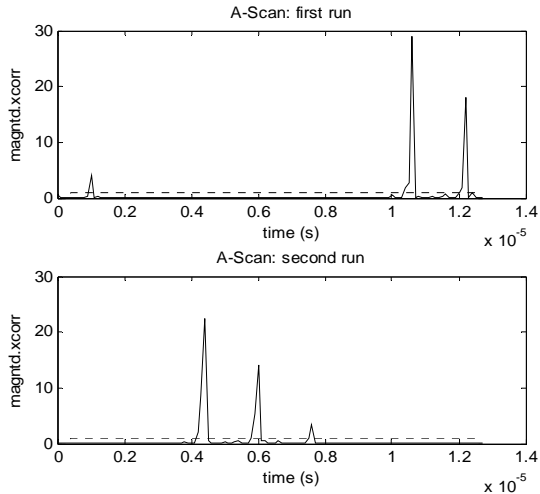

**Fig.1. Uncertain delay in USRP-based Radars: The peaks are shifted by uncertain amount of delays that makes their position change time by time with respect to reference window (dashed lines).**

Figure 1 shows actual measurement on USRP-based radar system. The received and cross-correlated signal is compared with generated code in the GRC block. Dashed line shows an observation window that is as a reference. In a synchronized system, both of the first run and the second one should be consistent. Uncertain delay makes them shifted by random delay.

On the other hand, in a single receiver radar system, the cross-correlation or timing comparison should be done for the received wave with the generated one in the GNU Radio processing block. Uncertain delay on the way from PC to USRP will affect the actual delay time measurement between target and the radar antenna.

Therefore, the range measurement will not give a correct value.

### 3.1 Proposed Solution
#### 3.2.1  Basic Ideas

The following pictures show the basic principle of the proposed solution. Figure 2 shows block diagram of USRP-based radar and signal paths. Modulated signal is transmitted by a USRP, then hitting the targets, and reflected back to the receiver. On the way to the antenna, the signal from host computer will suffered from random delay due to the communication protocol between the host computer and the USRP.

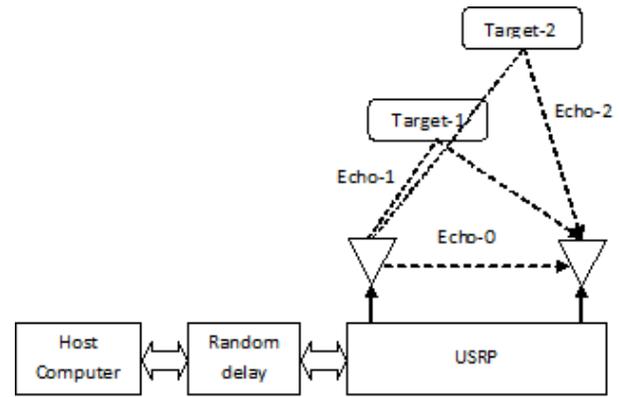

**Fig.2. Various delay in USRP-Based Radar: uncertain/random delay by communication protocol, direct reception/leaked signal, delay by target-1, and delay by target-2.**

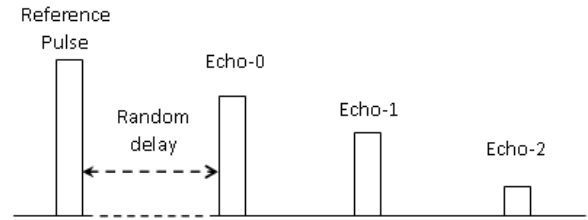

**Fig.3 Timing diagram of reference pulse and echoes**

The timing diagram of a SDR-radar ranging for two targets is shown in Fig.3. Uncertain delay causing random error when comparing Echo-1 and Echo-2 with the Tx-pulse. Actually, this uncertainty occurs between Tx-pulse and Echo-0, which is direct reception from Tx-to-Rx. When Echo-0 is used as a reference, determination of Echo-1 and Echo-2 will indicates the true values; therefore the measurement of the range will be correct.

*3.2 Evaluation Testbed: A Random Code SDR -Radar*

A coded radar detect target by cross-correlating transmitted code with the received one. The peak in the cross-correlation function indicates the target location.

Therefore, it will be consisting of: a random code generator, a modulator, a transmitter, (two) antenna(s), a receiver, a demodulator, and a correlator.

block is then followed by *Repeat* block so that the desired range bin resolution is achieved.

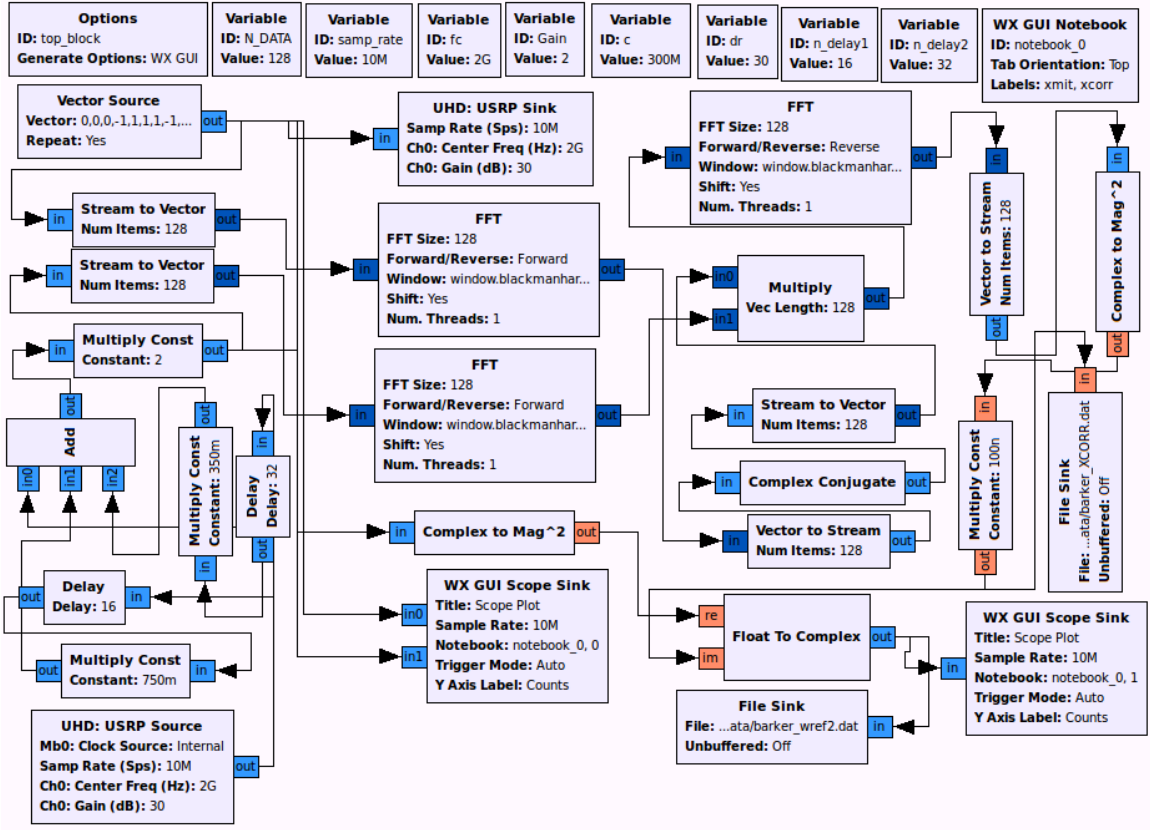

**Fig.4 GRC Block Diagram of the 128 Length Random Code Radar**

A simple version of this radar can be constructed from the following GRC (GNU-Radio Companion) blocks shown in Fig.4.

*3.2.1 Pseudo-Noise/Random Code Generator*

The longest Barker Code is 13, which normally is insufficient for a radar application. Therefore, we constructed a product of Barker Code length 11 with itself. The result is a 121 length composite Barker Code, then we pad it with 7 zeros to construct a 128 length random code (PNC: Pseudo Noise Code), considering the FFT employed in cross-correlation needs $2^N$ length input.

PNC128 = [0, 0, 0, B11⊗B11, 0, 0, 0, 0]  ; where

B11⊗B11=[+B11,+B11,+B11,-B11,-B11, -B11, +B11, -B11, -B11,+B11, -B11];   and
B11=[+1,+1,+1,−1,−1,−1,+1,−1,−1,+1,−1]

In the GRC, the generator is implemented as *Vector Source* block, and filling the parameter with PNC128. The

*3.2.2 Modulator, Transmitter, and Antenna(s)*

In the simplest implementation, translating the Barker Code into polarized truncated sine wave is not necessary. Therefore, the modulator an the transmitter in our case will be the *UHD-USRP Sink* block, set at desired frequency center, sufficient Gain factor, and TX antenna.

*3.2.3 Receiver and Demodulator*

In pair with the transmitter, *UHD-USRP Source* with identical parameter is configured as a receiver. Bandiwdth limitation will transform square wave of the code into a smooth ones.

*3.2.4 Correlator*

This block will perform cross-correlation of transmitted Code with the received (and demodulated) one. The GRC block implements transform-domain cross-correlation by the following stages:

- Both of the transmitted and recevied codes are Fourier Transformed (*forward FFT*)
- Take the complex conjugate of the transformed transmitted code

- Take the dot product (multiply) of them
- Inverse Fourier transform the product in the last stage (*reverse FFT*)

The result is a time-domain cross correlation of the sequences, with peaks at the target position. The largest peak is identified as direct reception. Since the cross-correlation is performed by taking 128 samples of received code, while the random delay shifts the code whose transmitted sequentially, the effect will be circular shift of the code in the received samples. Accordingly, re-ordering the peaks in the correlation function (A-scan) will be simply performed by detecting the highest one indicating direct reception, then circularly shift the correlation function so that the strongest peak moved to the first bin.

## 4. Experiments

In the experiment, we simulate the 128-length random code radar with sampling rate 10Msps. Assuming $c=3\times10^8$ *m/s*, the maximum unambiguous range will be 1,920 *m*, while range resolution is 15 *m*. There are two targets at range positions 240 *m* and 480 *m*, respectively. Random delay between PC and USRP is assumed distributed uniformly, shifting the echoes among 128 bins range. First, the system is simulated with Matlab. Two of instances of simulations are showed in the following Fig.5.

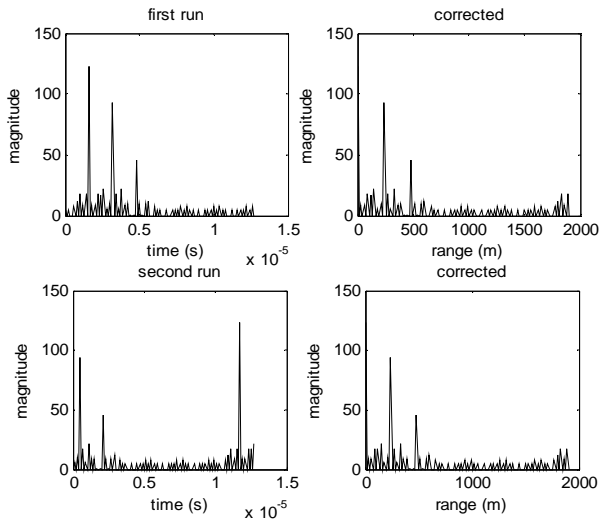

**Fig. 5 Matlab Simulation Results**

The first run gives a random delay of 0.16 *ms*, while the second one is 1.17 *ms*; shown in the left parts of the figures. The peaks are detected and the highest one indicating direct reception is then used as a reference. The A-scans are shifted circularly so that the reference moves into the first bin in corresponding scan. Right parts of the figures shows consistent results in both runs, where the targets are moved to the correct range bin.

Using the same parameters as before, an experiment is conducted using URSP. Since WBX daughterboard is available, we choose 2.2GHz carrier frequency. Actual transmission and reception is done, however the delay is simulated at the receiver side by two delay blocks, each corresponds with the targets. The output of the correlation is both displayed on a *Scope* and stored into a binary file. Figure 6 shows the result before and after circular shift. We observe the targets at corresponding bins correctly.

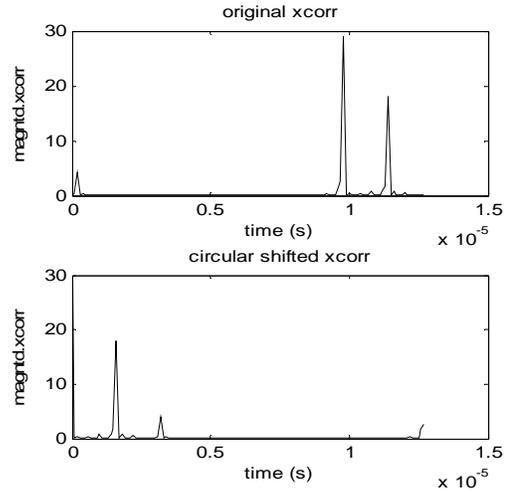

**Fig.6 Cross-Correlation Sequence Output of the SDR Radar Implemented on GRC and USRP2**

## 5. Conclusions

In this paper, we have shown that a simple method by choosing direct reception signals as a reference can be used to manage random delay problem in USRP-based radars. Correction stage is done by simply choose peak location, followed by circular shift. The workability of the method is demonstrated by Matlab simulation and USRP realization of 128-length random code radar.

**Acknowledgement**

This research is supported by a Grant from Ministry of Research and Technology, Republic of Indonesia, 2013.